\documentclass[twocolumn,
 amsmath,
 amssymb,
 aps,
 nofootinbib,
 preprintnumbers
]{revtex4-1}
\usepackage[utf8]{inputenc}
\usepackage{simplewick}
\usepackage{footnote}
\usepackage{graphicx}
\usepackage{dcolumn}
\usepackage{bm}
\usepackage{stmaryrd}
\usepackage{amsmath}
\usepackage{amssymb}
\usepackage{bbm}
\usepackage{amsfonts}
\usepackage{xcolor}
\usepackage{footmisc}
\usepackage{comment}
\usepackage{ulem}
\usepackage{slashed}
\usepackage{multirow}
\usepackage{appendix}
\usepackage{float}
\usepackage[linktocpage=true,bookmarksnumbered=true,colorlinks=true,plainpages,linkcolor=blue,citecolor=red]{hyperref}

\begin{document}
\preprint{ADP-23-06/T1215}
\title{Graviton-photon production with a massive spin-2 particle}

\author{Joshua A. Gill}
\author{Dipan~Sengupta}
\author{Anthony~G.~Williams}

\affiliation{ARC Centre of Excellence for Dark Matter Particle Physics, Department of Physics, University of Adelaide, South Australia 5005, Australia}
 
\begin{abstract}
    A recent letter \cite{Cai2022} within a phenomenological dark matter framework with a massive graviton in the external state argued that a divergence with increasing centre-of-momentum energy arising from the longitudinal polarizations of the graviton. In this letter we point out that in processes such as graviton-photon production from matter annihilation, $f\bar f\to G\gamma$, no such anomalous divergences occur at tree-level. This then applies to other tree-level amplitudes related by crossing symmetry such as $\gamma f \to G f$, $G f \to \gamma f$, $\gamma \bar f \to G \bar f$, $f\to f G \gamma$ and so on. We show this by explicitly computing the relevant tree-level diagrams, where we find that delicate cancellations ensure that all anomalously growing terms are well-regulated. Effectively at tree-level this is consistent with the operation of a Ward identity associated with the external photon for such amplitudes. The same tree-level results apply if the photon is replaced by a gluon. These results 
    have important consequences for model construction in areas including dark matter, gravity and high-energy physics.
\end{abstract}
 
\maketitle
 
\section{Introduction}
In the last few years, there has been a renewed interest in dark matter and phenomenological models with massive spin-2 particles \cite{Garny:2015sjg,Folgado:2019sgz,Cai2022,Kang:2020huh,deGiorgi:2021xvm,Bernal:2020ili}. While some of these approaches are simplified constructions of an underlying compact extra-dimensional theory \cite{Rueter:2017nbk}, others are effective field theories with a single massive graviton \cite{Kraml:2017atm}. In a number of these approaches, it has been shown that there are enhancements in matrix elements and cross-sections due to the longitudinal polarizations of the graviton, which grow like $\mathcal{O}(s/M_{G}^{2})$ at high energies, where $\sqrt{s}$ is the centre-of-momentum energy and $M_{G}$ the mass of the massive spin-2 particle \cite{Garny:2015sjg,Cai2022,Kraml:2017atm}. These require a lower bound on the graviton mass $M_{G}$ in order for the theory to be effective at large $s$\footnote{The $M_{G}\to 0$ limit is not smooth and leads to the famous vanDam-Veltman-Zakharov discontinuity \cite{vanDam:1970vg,Zakharov:1970cc}.}. This high energy scaling is expected in a naive Fierz-Pauli theory \cite{Fierz:1939ix} or extensions like bigravity/dRGT gravity \cite{deRham:2010kj,deRham:2014zqa,Hinterbichler:2011tt}\footnote{At high energies the scattering amplitudes of massive gravitons ($GG\to GG$) in the Fierz-Pauli theory grows as $s^{5}/(M_{G}^{8}M_{Pl}^{2})$, which can be estimated from power counting arguments \cite{Arkani-Hamed:2002bjr,Schwartz:2003vj,deRham:2014zqa,Hinterbichler:2011tt}. It can be shown that in extensions like dRGT gravity, this scaling can be improved to $s^{3}/(M_{G}^{4}M_{Pl}^{2})$ by adding higher order terms in the potential, but not beyond \cite{deRham:2010kj}.\looseness=-1}. However, in Kaluza-Klein (KK) theories with compact extra dimensions \cite{Han1999,Randall:1999ee,Randall:1999vf}, spin-2 KK mode scatterings are unitarized due to the underlying higher dimensional diffeomorphism invariance \cite{Chivukula:2020hvi,SekharChivukula:2019qih,SekharChivukula:2019yul,Hang:2021fmp,Hang:2022rjp}\footnote{It can be shown through a rigorous calculation that these cancellations persist even when the radial mode, the radion, gets a mass via the Goldberger-Wise mechanism \cite{Chivukula:2022tla,Chivukula:2021xod}}. In KK theories, these results have been extended to coupling with matter localized on the four-dimensional brane \cite{deGiorgi:2020qlg}. \looseness=-1

In a recent letter \cite{Cai2022}, the authors claimed that in a simple graviton-gluon production process, $f\bar{f}\to G g$, with a gluon and a massive spin-2 particle in the external state, there is a chiral-symmetry-breaking enhancement due to a massive on-shell external fermion. They
concluded that the squared matrix elements for the longitudinal polarizations grow proportional to $\left[\left(s/M_{Pl}^{2}\right) \left(m_{f}^{4}/M_{G}^{4}\right)\right]$ at high energies, implying an increase with increasing fermion mass $m_{f}$ and a very strong enhancement with decreasing graviton mass, $M_{G}\to 0$. 
This should be compared with a growth of the form $|\mathcal{M}|^{2}\propto\mathcal{O}(s/M_{Pl}^{2})<1$ for a massless graviton theory since theories involving gravitons are effective 
field theories which are valid for $s\ll M_{Pl}$.
In Ref.~\cite{Cai2022}
the resulting enhancement was then used to estimate the relic density in a freeze-in dark matter model with a cosmologically stable light KK graviton. This model then showed a dramatic enhancement in the velocity averaged cross-section for $M_{G} \ll m_{f}$. 

The result also implied that even in a compactified extra-dimensional setup with massive KK modes in the external state, this enhancement should persist even when the full KK spectrum is taken into account, since there is no cancellation mechanism, in contrast to expectations of scaling of KK graviton scatterings respecting higher dimensional gauge/diffeomorphism invariance. 

Such an enhancement would also have
significant phenomenological consequences for the production of KK gravitons at high-energy colliders within extra-dimensional models, which would predict anomalously growing cross-sections for fermion-initiated processes, in conflict with previous estimates 
performed
in the massless case \cite{Giudice1999}. Furthermore,
such a scenario would also mean that even with extremely feeble couplings, this process 
could be detectable in
direct detection experiments.
\looseness=-1
 Phenomenological calculations with a massive spin-2 particle in the external state or as a portal to a  dark sector have been plagued with issues such as low scale unitarity violation and infrared cut-offs induced by
 a massive graviton or a massive spin-2 KK particle.  
 Such calculations
 range from relic density calculations \cite{Cai2022,Garny:2015sjg,Bernal:2018qlk,Bernal:2020fvw,Kang:2020huh,deGiorgi:2021xvm} to direct detection estimates \cite{Carrillo-Monteverde:2018phy}.
It is important to understand from fundamental principles whether these calculations are consistent from a field theory perspective.
The results presented here provide a simple first check on all such calculations and indeed our results have already proved of use\footnote{Following our paper references \cite{Voronchikhin:2023znz,Voronchikhin:2022rwc} and \cite{Jodlowski:2023yne}  have appeared with applications in direct detection experiments and forward physics experiments that confirm our results.}

In this letter, we explicitly calculate the graviton-photon production process\footnote{Graviton photoproduction $\gamma f\to G f$ has been calculated previously in \cite{Bjerrum-Bohr:2014lea} for massless gravitons.}, 
\begin{equation}
    f\bar{f}\to G \gamma,
    \label{Eq:fd1}
\end{equation}
where $G$ represents a massive spin-2 particle, and $\gamma$ the massless on-shell photon.  We show that the full tree-level squared amplitude at high energies grows as $|\mathcal{M}|^{2}\propto\mathcal{O}(s/M_{Pl}^{2})$, and there are no terms proportional to $m_{f}^{4}/M_{G}^{4}$, implying no enhancements or divergences as $M_{G}\to 0$ for finite fermion masses, contrary to the suggestion in \cite{Cai2022}. We demonstrate that although individual terms in the  $s,~t,~u$ and contact interactions grow as $\mathcal{O}(1/M_{G}^{2})$, due to the longitudinal polarizations of the massive graviton, delicate cancellations at tree-level ensure that the full amplitude has no low energy cut-off, for all incoming helicities of the fermion and outgoing helicities of the massless photon and the massive graviton.  An identical scaling of amplitudes at high energies is observed if a gluon replaces the photon. The only difference is replacing the electromagnetic coupling with the strong coupling. 
In what follows, we detail the calculation and present the full amplitude as a function of the centre-of-momentum energy $\sqrt{s}$ and scattering angle $\theta$.\looseness=-1
 
\section{Framework and Formalism}

We use the `mostly minus' metric convention for the flat four-dimensional Minkowski spacetime background (4D) $\rm \eta_{\mu\nu}\equiv\rm Diag(+1,-1,-1,-1)$, which is also used to raise and lower indices. Metric fluctuations $h_{\mu\nu}(x)$\footnote{From here on we will drop the spacetime index $x$, and in momentum space $k$ unless explicitly specified.} around the flat Minkowski background is expressed as,
\begin{equation}
    \eta_{\mu\nu}\to \eta_{\mu\nu}+\kappa h_{\mu\nu}(x)\equiv \tilde{G}_{\mu\nu}(x),
\end{equation}
 which define the spin-2 graviton in 4D. The dimensionfull coupling $\kappa$ is related to the fundamental 4D Planck mass as $\kappa=2/M_{Pl}=\sqrt{16\pi G_{N}}$.
 A theory of massive gravity, dubbed as the Fierz-Pauli theory, can be expressed as, 
\begin{equation}
    \mathcal{L}= \frac{M_{Pl}^{2}}{2}\sqrt{-|\tilde{G}|}R + \frac{M_{G}^{2}}{2}(h^{2} -h_{\mu\nu}^{2}). 
\end{equation}
Here $|\tilde{G}|$ is the determinant of 4D metric with fluctuations and $h \equiv \eta^{\mu\nu}h_{\mu\nu}$. The first term represents the Einstein-Hilbert piece, $R$ being the Ricci scalar, while the second represents the Fierz-Pauli mass term. In theories of compact extra dimensions, the same mass terms for spin-2 KK gravitons appear after compactification, along with the massless graviton. For example, in Randall-Sundrum models in warped extra dimensions, the masses of the $n^{th}$ modes of the  spin-2 KK gravitons are given by (in the large curvature limit) $m_{n} \simeq x_{n}k e^{-\pi kr_{c}}$, where $x_{n}$ are the zeros of the Bessel function of the first kind, $k$ is the curvature and $r_{c}$ the radius of the compactification.

\begin{figure}[H]
    \centering
    \includegraphics[width=.22\textwidth]{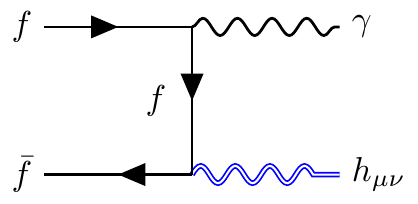}
    \includegraphics[width=.22\textwidth]{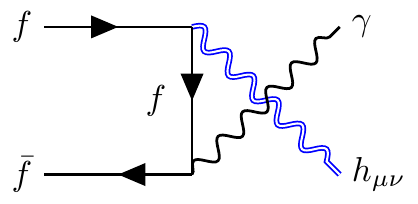}
    \includegraphics[width=.24\textwidth]{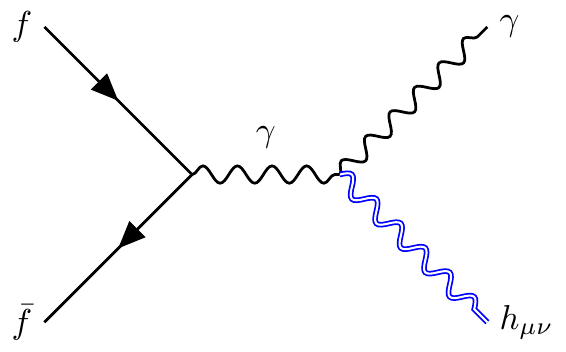}
    \includegraphics[width=.19\textwidth]{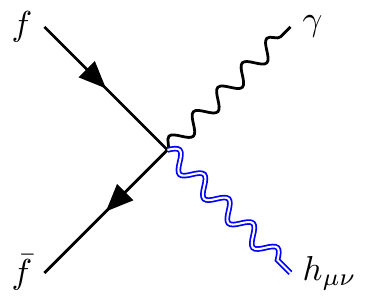}
\caption{Feynman diagrams for the process $f (p_{1})+\bar{f}(p_{2})\to \gamma (k_{1})+ G(k_{2})$}
\label{fd1}
\end{figure}

The couplings of the graviton to matter (scalars, fermions or vectors) can be expressed by the following action,
\begin{equation}
    \mathcal{S}_{M}= \int d^{4}x~ \mathcal{L}(\tilde{G},s,v,f),
\end{equation}
which upon expanding to order $\kappa$ in the metric fluctuation yields, 
\begin{equation}
    \mathcal{S}_{M}= -\frac{\kappa}{2}\int d^{4}x~h_{\mu\nu}T^{\mu\nu}(s,v,f).
\end{equation}
The stress energy tensor $T_{\mu\nu}$ is given by, 
\begin{equation}
    T_{\mu\nu} = \left( -\eta_{\mu\nu}\mathcal{L} + 2\frac{ \delta \mathcal{L}}{\delta \tilde{G}^{\mu\nu}} \right)|_{\tilde{G}=\eta}.
\end{equation}
For fermions, the stress-energy tensor must be calculated using the Vielbein formalism as performed in \cite{Han1999,Giudice1999}. We follow \cite{Han1999,Giudice1999} for the conventions and Feynman rules.  The process of interest here is graviton-photon production via the annihilation of a fermion and anti-fermion pair, as expressed in Eq. \ref{Eq:fd1}. 
The four diagrams shown in Fig. \ref{fd1}, $t$-, $u$-, $s$-channels and a contact term, respectively, are the only tree-level interactions.
The vertex rules are derived in \cite{Han1999} and are listed in the supplementary material. The coupling between the fermion and the photon is $g_{f}e$, where $e\equiv |e|$ is the magnitude of the charge of the electron. \looseness=-1

We define the following variable, which will appear in the $s$-channel diagram with a gauge parameter $\xi$, as:
\begin{equation}
    \begin{split}
        & W_{\mu\nu\alpha\beta}\left(k_{1}, k_{2}; \xi\right) 
           = (1/2)\eta_{\mu\nu}\left(k_{1\beta}k_{2\alpha} - k_{1}\cdot k_{2}\eta_{\alpha\beta}\right) \\
        & \qquad\quad + \eta_{\mu\alpha}\left(k_{1}\cdot k_{2}\eta_{\nu\beta} - k_{1\beta}k_{2\nu}\right) \\
        & \qquad\quad + \eta_{\alpha\beta}k_{1\mu}k_{2\nu} - \eta_{\mu\beta}k_{1\nu}k_{2\alpha} \\
        & \qquad\quad - (1/\xi)\left\{\left(\eta_{\nu\beta}k_{1\mu}k_{1\alpha} + \eta_{\nu\alpha}k_{2\mu}k_{2\beta}\right) \right. \\
        & \qquad\quad \left. - (1/2)\eta_{\mu\nu}\left(k_{1\alpha}k_{1\beta} + k_{2\alpha}k_{2\beta} + k_{1\alpha}k_{2\beta}\right)\right\}\label{WDef}.
    \end{split}
\end{equation}
The photon propagator is defined as,
\begin{align}
    \Delta_{\mu\nu}\left(Q\right) = -\dfrac{i}{Q^{2}}\left[\eta_{\mu\nu} + \left(\xi - 1\right)\dfrac{Q_{\mu}Q_{\nu}}{Q^{2}}\right]. 
\end{align}
For simplicity, we work in Feynman gauge $\xi = 1$. The fermion propagator with momentum $Q$ and mass $m_{f}$ travelling in the direction of the fermion flow is given by,
\begin{align}
    S_{F}\left(Q\right) & = \dfrac{i\left(\slashed{Q} + m_{f}\right)}{Q^{2} - m_{f}^{2}}. 
\end{align}
We define the Mandelstam variables such that,
\begin{align}
    s & = \left(p_{1} + p_{2}\right)^{2} = \left(k_{1} + k_{2}\right)^{2}, \\
    t & = \left(p_{1} - k_{1}\right)^{2} = \left(p_{2} - k_{2}\right)^{2}, \\
    u & = \left(p_{1} - k_{2}\right)^{2} = \left(p_{2} - k_{1}\right)^{2}.
\end{align}
Choosing the $\hat{z}$ direction as the centre-of-momentum frame, with an outgoing massless photon and a massive graviton with mass $M_{G}$, we can express the four-momenta of various particles as,
\begin{align}
    & p_{1}^{\mu} = \left(E_{p_{1}}, \, \left|\boldsymbol{p}\right|\hat{z}\right), \qquad\quad p_{1}^{2} = m_{f}^{2}, \\
    & p_{2}^{\mu} = \left(E_{p_{2}}, \, -\left|\boldsymbol{p}\right|\hat{z}\right), \qquad p_{2}^{2} = m_{f}^{2}, \\
    & k_{1}^{\mu} = E_{k_{1}}\left(1, \, -\hat{k}\right), \qquad\,\, k_{1}^{2} = 0, \\
    & k_{2}^{\mu} = \left(E_{k_{2}}, \, \boldsymbol{k}\right), \qquad\qquad\, k_{2}^{2} = M_{G}^{2}.
\end{align}
 The momentum ($\boldsymbol{k}$) of the outgoing graviton and photon  are given in terms of the inclination and azimuthal angle pairing $\left(\theta,\,\phi\right)$  as
  $\boldsymbol{k} = \left|\boldsymbol{k}\right|\left(s_{\theta}c_{\phi}, \,s_{\theta}s_{\phi},\,c_{\theta}\right)$, where $c_{\theta }\equiv \cos\theta$ and $s_{\theta} \equiv \sin\theta$.
The polarizations for the external on-shell photon are defined in the usual way, 
\begin{align}
    \varepsilon_{\pm1}^{\mu}\left(k_{1}\right) & = \pm\dfrac{e^{\pm i\phi}}{\sqrt{2}}\bigg(0, -c_{\theta}c_{\phi}\pm is_{\phi},-c_{\theta}s_{\phi}\mp ic_{\phi},s_{\theta}\bigg) \label{polPM1}. 
 \end{align}   
A helicity-$\lambda_{G}$ massive graviton carries five polarizations $\varepsilon^{\mu\nu}_{\lambda_{G}}(k)$. These are grouped into two transverse, and three longitudinal polarizations, which can be 
split into two helicity-1 modes and one helicity-0 mode, defined  respectively as \cite{Chivukula:2020hvi}\looseness=-1, 
\begin{align}
    \lambda_{G} = \pm2, \,\, \varepsilon_{\pm2}^{\mu\nu} & = \varepsilon_{\pm1}^{\mu}\varepsilon_{\pm1}^{\nu}, \\
    \lambda_{G} = \pm1, \,\, \varepsilon_{\pm1}^{\mu\nu} & = \dfrac{1}{\sqrt{2}}\bigg[\varepsilon^{\mu}_{\pm1}\varepsilon^{\nu}_{0} + \varepsilon^{\mu}_{0}\varepsilon^{\nu}_{\pm1}\bigg], \\
    \lambda_{G} = 0, \quad\, \varepsilon_{0}^{\mu\nu} & = \dfrac{1}{\sqrt{6}}\bigg[\varepsilon_{+1}^{\mu}\varepsilon_{-1}^{\nu} + \varepsilon_{-1}^{\mu}\varepsilon_{+1}^{\nu} + 2\varepsilon_{0}^{\mu}\varepsilon_{0}^{\nu}\bigg],
\end{align}
where $\varepsilon_{\pm1}^{\mu}$ are the usual polarization vectors for the photon defined in Eq. \ref{polPM1}, while the helicity-0 polarization is defined by, 
\begin{align}    
    \varepsilon_{0}^{\mu}\left(k_{2}\right) & = \dfrac{E_{k_{2}}}{M_{G}}\bigg(\sqrt{1 - \dfrac{M_{G}^{2}}{E_{k_{2}}^{2}}}, \, \hat{k}\bigg). \label{pol0}
\end{align}
The polarization vectors for momentum $k_{2}$ are defined using the same angle pairs $\left(\theta, \phi\right)$. Without loss of generality, we have chosen $\phi = 0$ in the calculation. 

Choosing the centre-of-momentum frame for the incoming particles with four-vectors $p_{1}$ and $p_{2}$, and outgoing four-vectors $k_{1}$ and $k_{2}$, the outgoing energies $E_{k_{1}}$ and $E_{k_{2}}$ can be expressed in terms of the Mandelstam variable $s$ and the mass of the graviton $M_{G}$ as, 
\begin{align}
    E_{k_{1}} = \dfrac{s - M_{G}^{2}}{2\sqrt{s}}, \qquad E_{k_{2}} = \dfrac{s + M_{G}^{2}}{2\sqrt{s}}. 
\end{align}

For the Feynman diagrams depicted in Fig. \ref{fd1}, with an incoming fermion $f(p_{1})$ and anti-fermion $\bar{f}(p_{2})$ scattering to a photon with polarization $\varepsilon_{\lambda}(k_{1})$ and a massive graviton with polarization $\varepsilon_{\lambda_{G}}^{\mu\nu}$, the matrix elements\footnote{The matrix elements of above disagree with \cite{Cai2022} in the $u$-channel only of Eq. (\ref{uchannel}). This can be a potential cause of the differing results.\looseness=-1} for the $t,~u,~s$ and the contact diagrams are respectively given by,  
\begin{align}
    \begin{split}
        \mathcal{M}_{t} & = -\dfrac{\kappa g_{f}e}{8} \bar{v}_{\lambda_{1}}\left(p_{2}\right)\big[\gamma_{\mu}P_{\nu} \!+\! \gamma_{\nu}P_{\mu} \!-\! 2\eta_{\mu\nu}\left(\slashed{P} \!-\! 2m_{f}\right)\big]\\
        \times & \left(\dfrac{\slashed{p}_{1} - \slashed{k}_{1} + m_{f}}{t - m_{f}^{2}}\right)\slashed{\varepsilon}^{*}_{\lambda_{\gamma}}\left(k_{1}\right)\varepsilon_{\lambda_{G}}^{*\mu\nu}\left(k_{2}\right)u_{\lambda_{2}}\left(p_{1}\right),
        \label{tchannel}
    \end{split} \\
    \begin{split}
        \mathcal{M}_{u} & = -\dfrac{\kappa g_{f}e}{8} \bar{v}_{\lambda_{1}}\left(p_{2}\right)\slashed{\varepsilon}^{*}_{\lambda_{\gamma}}\left(k_{1}\right)\left(\dfrac{\slashed{p}_{1} - \slashed{k}_{2} + m_{f}}{u - m_{f}^{2}}\right) \\
        \times & \big[\gamma_{\mu}K_{\nu} \!+\! \gamma_{\nu}K_{\mu} \!-\! 2\eta_{\mu\nu}\left(\slashed{K} \!-\! 2m_{f}\right)\big]\varepsilon_{\lambda_{G}}^{*\mu\nu}\left(k_{2}\right)u_{\lambda_{2}}\left(p_{1}\right), \label{uchannel}
    \end{split} \\
    \begin{split}
        \mathcal{M}_{s} & = \dfrac{\kappa g_{f}e}{2s}\bar{v}_{\lambda_{1}}\left(p_{2}\right)\gamma^{\alpha}\big[W_{\mu\nu\alpha\beta}\left(Q, k_{1};\xi\right) \\
        + & W_{\nu\mu\alpha\beta}\left(Q, k_{1},\xi\right)\big]\varepsilon^{*\beta}_{\lambda_{\gamma}}\left(k_{1}\right)\varepsilon_{\lambda_{G}}^{*\mu\nu}\left(k_{2}\right)u_{\lambda_{2}}\left(p_{1}\right),
        \label{schannel}
    \end{split} \\
    \begin{split}
        \mathcal{M}_{c} & = \dfrac{\kappa g_{f}e}{4}\bar{v}_{\lambda_{1}}\left(p_{2}\right)\big[\gamma_{\mu}\eta_{\nu\alpha} + \gamma_{\nu}\eta_{\mu\alpha} - 2\eta_{\mu\nu}\gamma_{\alpha}\big] \\
        \times & \varepsilon^{*\alpha}_{\lambda_{\gamma}}\left(k_{1}\right)\varepsilon_{\lambda_{G}}^{*\mu\nu}\left(k_{2}\right)u_{\lambda_{2}}\left(p_{1}\right). \label{cchannel}
    \end{split}
\end{align}
Here, $ P \equiv \left(p_{1} - k_{1} - p_{2}\right) = \left(k_{2} - 2p_{2}\right)$, 
$Q \equiv -\left(p_{1} + p_{2}\right)$ and $K \equiv \left(p_{1} + k_{1} - p_{2}\right) = \left(2p_{1} - k_{2}\right)$. 
The fermions have spin states states $\lambda_{1}, \lambda_{2} = \uparrow \text{or}\downarrow$, and the photon has polarization states $\lambda_{\gamma} = \pm1$. The graviton has polarization states $\lambda_{G} = \pm 2,\, \pm 1$ and $0$.

\section{Results and Conclusion}

We first note that there are 40 combinations of outgoing helicities, with the corresponding incoming states of the spinors, with 16 each in helicity-2 and helicity-1 modes supplemented by 8 in helicity-0 modes. The helicity-2 modes correspond to polarizations of the massless graviton and have no bad small-mass behaviour. The helicity-0 mode exhibits the worst growth with decreasing graviton mass due to two factors of $\varepsilon^{\mu}_{0}$, where each factor grows as $\mathcal{O}(1/M_{G})$. The total matrix element is the sum of $s,~t,~u$ and contact diagrams, which we, therefore, expand as a series in the mass of the graviton $M_{G}$ to analyze if there are any divergences in the massless limit $M_{G}\rightarrow0$,
\begin{equation}
\mathcal{M}(s,\theta)= \sum_{\sigma\in\mathbb{Z}} M_{G}^{\sigma}\mathcal{M}(\theta).\label{PowerSeriesDef}
\end{equation}

The entire matrix element for these diagrams is non-trivial, and thus Mathematica \cite{Mathematica} was employed to compute the matrix element for each polarization symbolically.
We also observe that several polarization combinations vanish simply by helicity conservation and selection rules, as tabulated in the supplementary materials.\looseness=-1
\begin{table}[t]
\begin{tabular}{|p{1cm}||p{7cm}|}
    \hline
    \multicolumn{2}{|c|}{Helicity-0 External Graviton: $\left(u, \bar{v}, \gamma, G\right) = \left(\uparrow,\uparrow,+1,0\right)$} \\
    \hline
    \hline
    \multicolumn{2}{|c|}{Coefficient: $s\left(\kappa g_{f}e /4\sqrt{3}\right)\left(m_{f}/M_{G}^{2}\right)\sin\theta$} \\
    \hline
    \hline
    $\mathcal{M}_{t}$ & $1 + \cos\theta\sqrt{1 - 4m_{f}^{2}/s}$ \\
    \hline 
    $\mathcal{M}_{u}$ & $1 - \cos\theta\sqrt{1 - 4m_{f}^{2}/s}$\\
    \hline
    $\mathcal{M}_{s}$ & $-2$ \\
    \hline 
    $\mathcal{M}_{c}$ & $0$\\
    \hline
    \hline
    $\sum\mathcal{M}$ & 0\\
    \hline
\end{tabular}
\caption{The cancellations for a helicity-0 external graviton are presented. 
Note that $\mathcal{M}_{t,u,s,c}$ represent the matrix element contributions for the diagrams depicted in Fig. \ref{fd1} }
\label{tab:hel0}
\end{table}

Suppose that we choose some polarization state to investigate and interrogate the results from each Feynman diagram to determine the origin of the divergence.

The leading divergent terms for the longitudinal polarization mode $\left(u, \bar{v}, \gamma, G\right) = \left(\uparrow,\uparrow,+1,0\right)$, are demonstrated in Table \ref{tab:hel0} for the $s,~t,~u$ and contact diagrams.
We notice that while each of the $s,~t~,u$ diagrams grow proportional to ($m_{f}/M_{G}^{2}$), as expected from power counting arguments, the sum vanishes identically, leading to regular behaviour in the limit as $M_{G}\to 0$. 

Scanning through every possible combination of the helicities, we find that the divergences in each channel exactly cancel when all channels are summed.
\footnote{It is reasonable to question if the cancellations occur when the amplitude is squared  since mixing terms between diagrams become relevant. We have checked that this holds by virtue of L'Hôpital's rule.}
 
Therefore, the leading order term in the limit as $M_{G}\to0$ for all polarization combinations is a constant, including the scalar, vector and longitudinal polarizations of the graviton. Thus, no divergences persists once all diagrams are summed and hence, $\sigma \ge 0$ in Eq. (\ref{PowerSeriesDef}).

Squaring the amplitude we find no divergences in the limit as the graviton becomes massless $M_{G}\to0$. The leading order in the limit is a constant term with respect to the graviton mass $M_{G}$,

\begin{equation}
    \lim_{M_{G}\rightarrow0}\left|\mathcal{M}\left(s,\theta\right)\right|^{2} = \mathcal{O}\left(M_{G}^{0}\right).
\end{equation}

Considering now the high energy limit with a finite graviton mass $M_{G}$, the leading high energy contribution to the matrix element for the helicity-0 modes is proportional to the fermion mass and is given by\footnote{Similar cancellations occur for helicity-1 modes and is documented in the supplementary materials.},

\begin{equation}
    \begin{split}
\lim_{s\to\infty}\sum_{\lambda_{G}=0}\left|\mathcal{M}\left(s,\theta\right)\right| & = \dfrac{2\kappa g_{f}e}{\sqrt{3}}m_{f}\csc\theta + \mathcal{O}\left(s^{-1}\right).
    \end{split}
\end{equation}

The series expansion in the high energy limit $\sqrt{s}\rightarrow\infty$ is a physically interesting one. For example, we observe no anomalous behaviour in the high energy limit for the unpolarized process\footnote{In \cite{Giudice1999}, the cross-section for $f\bar{f}\to \gamma~G_{KK_{m}}$ in the $m_{f}\to 0$ limit is provided, showing no enhancements proportional to $1/M_{KK}^{2}$. We agree with this result.}, 
\begin{align}
        & \lim_{s\rightarrow\infty}\sum_{\text{all spins}}\left|\mathcal{M}\left(s,\theta\right)\right|^{2}  = \dfrac{\left(\kappa g_{f}e\right)^{2}}{24}\Bigg\{6s\left[3 + \cos\left(2\theta\right)\right]  \\
        & \qquad\qquad
        + \bigg[27M_{G}^{2} - 14m_{f}^{2} - 6\left[M_{G}^{2} + 12m_{f}^{2}\right]\cos\left(2\theta\right) \nonumber \\
        & \qquad\qquad + 3\left[M_{G}^{2} + 2m_{f}^{2}\right]\cos\left(4\theta\right)\bigg]\csc^{2}\theta  + \mathcal{O}\left(s^{-1}\right)\Bigg\}. \nonumber
\end{align}

We next attempt to understand if there are underlying symmetry arguments that enforce the cancellation in terms proportional to powers of $1/M_{G}$.
Pathologies in massive gravity theories come primarily from  massive internal graviton propagators \cite{deRham:2010kj}. Since we do not have them at tree-level for the process of interest,
it is interesting to contemplate whether some QED-like Ward identity might effectively survive
in this situation. The inclusion of an external graviton source in QED does not alter the global $U(1)$ symmetry and so a conserved current will result. It seems reasonable to anticipate 
that an effective QED Ward identity might emerge in a careful treatment\footnote{For a derivation of Ward identity, see for example \cite{Williams__2022}.}. We have directly verified for our amplitudes that we do have an effective QED Ward identity operating since we find
\begin{equation}
     k_{1}^{\alpha}\mathcal{M}_{\alpha} = 0, 
    \end{equation}
where the quantities $\mathcal{M}_{\alpha}$ and $\mathcal{M}_{\mu\nu}$ are defined such that $\mathcal{M} \equiv \mathcal{M}_{\mu\nu\alpha}\varepsilon^{\alpha}\left(k_{1}\right)\varepsilon^{\mu\nu}\left(k_{2}\right) = \mathcal{M}_{\alpha}\varepsilon^{\alpha}\left(k_{1}\right) = \mathcal{M}_{\mu\nu}\varepsilon^{\mu\nu}\left(k_{2}\right)$. Furthermore we also anticipate that as long as the energy-momentum tensor $T_{\mu\nu}$ is conserved, an analogous identity should hold on the gravitational side, which we verify explicitly, 
\begin{equation}
     k_{2}^{\mu}k_{2}^{\nu}\mathcal{M}_{\mu\nu} = 0.
\end{equation}
This ensures that all contributions that grow as powers of $1/M_{G}$ in individual diagrams cancel out for any given process.

While we have only explicitly calculated for the case of $f\bar f\to G \gamma$ the above results
will also apply to other tree-level amplitudes related by crossing symmetry such as
    $\gamma f \to G f$, $G f \to \gamma f$, $\gamma \bar f \to G \bar f$,
    $f\to f G \gamma$ and so on.

The above results will also hold when a gluon replaces the photon in the external leg. For example, we note that this breaks down when considering two massive graviton emissions, i.e., a process like $f\bar{f}\to GG$, due to the presence of an $s$-channel 
diagram with a massive graviton in the internal propagator. In this case, there is no mechanism by which this cancellation can take place for a theory of massive gravity \cite{Falkowski:2020mjq}.

Therefore, we have demonstrated no enhancements in the limit as $M_{G}\to0$
in the matrix elements of massive graviton-photon scattering with initial fermion states, regardless of whether the fermion is massive or not, contrary to claims in \cite{Cai2022}. Hence, the dark matter
scenario for which the authors claim large enhancements in the velocity-averaged cross-section appears inconsistent with our calculation. Finally, we note that our calculation goes beyond correcting Ref.~\cite{Cai2022}. It illustrates
the importance of exploiting the 
deep theoretical field theoretic structures of massive gravity and extra dimensions as a check on calculations that are often very complex and difficult.
Our demonstration of an effective Ward identity at tree level is one very important example. Our calculations have significant phenomenological implications in the calculation of KK graviton production at the LHC, direct detection of KK graviton dark matter, as well as relic density calculation in spin-2 dark matter models.

{\bf{\underline{Acknowledgements}}}
JAG acknowledges the support he has
received for his research through the provision of an Australian Government Research Training Program Scholarship.
Support for this work was provided by the University of Adelaide and the Australian Research Council through the Centre of Excellence for Dark Matter Particle Physics (CE200100008). DS acknowledges the Mainz Institute of Theoretical Physics workshop `Towards the Next Fundamental Scale of Nature: New Approaches in Particle Physics and Cosmology", where this project originated. DS and JAG thank Seung J. Lee and Giacomo Cacciapaglia for illuminating conversations. DS also thanks R. Sekhar Chivukula, Xing Wang and Kirtimaan Mohan for discussions.

\bibliographystyle{utphys.bst}
\bibliography{reference}

\providecommand{\href}[2]{#2}\begingroup\raggedright\begin{thebibliography}{10}

\bibitem{Cai2022}
H.~Cai, G.~Cacciapaglia, and S.~J. Lee, ``Massive gravitons as feebly
  interacting dark matter candidates,''
  \href{http://dx.doi.org/10.1103/PhysRevLett.128.081806}{{\em Phys. Rev.
  Lett.} {\bfseries 128} (Feb, 2022) 081806}.
  \url{https://link.aps.org/doi/10.1103/PhysRevLett.128.081806}.

\bibitem{Garny:2015sjg}
M.~Garny, M.~Sandora, and M.~S. Sloth, ``{Planckian Interacting Massive
  Particles as Dark Matter},''
  \href{http://dx.doi.org/10.1103/PhysRevLett.116.101302}{{\em Phys. Rev.
  Lett.} {\bfseries 116} no.~10, (2016) 101302},
  \href{http://arxiv.org/abs/1511.03278}{{\ttfamily arXiv:1511.03278
  [hep-ph]}}.

\bibitem{Folgado:2019sgz}
M.~G. Folgado, A.~Donini, and N.~Rius, ``{Gravity-mediated Scalar Dark Matter
  in Warped Extra-Dimensions},''
  \href{http://arxiv.org/abs/1907.04340}{{\ttfamily arXiv:1907.04340
  [hep-ph]}}. [Erratum: JHEP 02, 129 (2022)].

\bibitem{Kang:2020huh}
Y.-J. Kang and H.~M. Lee, ``{Lightening Gravity-Mediated Dark Matter},''
  \href{http://dx.doi.org/10.1140/epjc/s10052-020-8153-x}{{\em Eur. Phys. J. C}
  {\bfseries 80} no.~7, (2020) 602},
  \href{http://arxiv.org/abs/2001.04868}{{\ttfamily arXiv:2001.04868
  [hep-ph]}}.

\bibitem{deGiorgi:2021xvm}
A.~de~Giorgi and S.~Vogl, ``{Dark matter interacting via a massive spin-2
  mediator in warped extra-dimensions},''
  \href{http://dx.doi.org/10.1007/JHEP11(2021)036}{{\em JHEP} {\bfseries 11}
  (2021) 036}, \href{http://arxiv.org/abs/2105.06794}{{\ttfamily
  arXiv:2105.06794 [hep-ph]}}.

\bibitem{Bernal:2020ili}
N.~Bernal and O.~Zapata, ``{Gravitational dark matter production: primordial
  black holes and UV freeze-in},''
  \href{http://dx.doi.org/10.1016/j.physletb.2021.136129}{{\em Phys. Lett. B}
  {\bfseries 815} (2021) 136129},
  \href{http://arxiv.org/abs/2011.02510}{{\ttfamily arXiv:2011.02510
  [hep-ph]}}.

\bibitem{Rueter:2017nbk}
T.~D. Rueter, T.~G. Rizzo, and J.~L. Hewett, ``{Gravity-Mediated Dark Matter
  Annihilation in the Randall-Sundrum Model},''
  \href{http://dx.doi.org/10.1007/JHEP10(2017)094}{{\em JHEP} {\bfseries 10}
  (2017) 094}, \href{http://arxiv.org/abs/1706.07540}{{\ttfamily
  arXiv:1706.07540 [hep-ph]}}.

\bibitem{Kraml:2017atm}
S.~Kraml, U.~Laa, K.~Mawatari, and K.~Yamashita, ``{Simplified dark matter
  models with a spin-2 mediator at the LHC},''
  \href{http://dx.doi.org/10.1140/epjc/s10052-017-4871-0}{{\em Eur. Phys. J. C}
  {\bfseries 77} no.~5, (2017) 326},
  \href{http://arxiv.org/abs/1701.07008}{{\ttfamily arXiv:1701.07008
  [hep-ph]}}.

\bibitem{vanDam:1970vg}
H.~van Dam and M.~J.~G. Veltman, ``{Massive and massless Yang-Mills and
  gravitational fields},''
  \href{http://dx.doi.org/10.1016/0550-3213(70)90416-5}{{\em Nucl. Phys. B}
  {\bfseries 22} (1970) 397--411}.

\bibitem{Zakharov:1970cc}
V.~I. Zakharov, ``{Linearized gravitation theory and the graviton mass},'' {\em
  JETP Lett.} {\bfseries 12} (1970) 312.

\bibitem{Fierz:1939ix}
M.~Fierz and W.~Pauli, ``{On relativistic wave equations for particles of
  arbitrary spin in an electromagnetic field},''
  \href{http://dx.doi.org/10.1098/rspa.1939.0140}{{\em Proc. Roy. Soc. Lond. A}
  {\bfseries 173} (1939) 211--232}.

\bibitem{deRham:2010kj}
C.~de~Rham, G.~Gabadadze, and A.~J. Tolley, ``{Resummation of Massive
  Gravity},'' \href{http://dx.doi.org/10.1103/PhysRevLett.106.231101}{{\em
  Phys. Rev. Lett.} {\bfseries 106} (2011) 231101},
  \href{http://arxiv.org/abs/1011.1232}{{\ttfamily arXiv:1011.1232 [hep-th]}}.

\bibitem{deRham:2014zqa}
C.~de~Rham, ``{Massive Gravity},''
  \href{http://dx.doi.org/10.12942/lrr-2014-7}{{\em Living Rev. Rel.}
  {\bfseries 17} (2014) 7}, \href{http://arxiv.org/abs/1401.4173}{{\ttfamily
  arXiv:1401.4173 [hep-th]}}.

\bibitem{Hinterbichler:2011tt}
K.~Hinterbichler, ``{Theoretical Aspects of Massive Gravity},''
  \href{http://dx.doi.org/10.1103/RevModPhys.84.671}{{\em Rev. Mod. Phys.}
  {\bfseries 84} (2012) 671--710},
  \href{http://arxiv.org/abs/1105.3735}{{\ttfamily arXiv:1105.3735 [hep-th]}}.

\bibitem{Arkani-Hamed:2002bjr}
N.~Arkani-Hamed, H.~Georgi, and M.~D. Schwartz, ``{Effective field theory for
  massive gravitons and gravity in theory space},''
  \href{http://dx.doi.org/10.1016/S0003-4916(03)00068-X}{{\em Annals Phys.}
  {\bfseries 305} (2003) 96--118},
  \href{http://arxiv.org/abs/hep-th/0210184}{{\ttfamily arXiv:hep-th/0210184}}.

\bibitem{Schwartz:2003vj}
M.~D. Schwartz, ``{Constructing gravitational dimensions},''
  \href{http://dx.doi.org/10.1103/PhysRevD.68.024029}{{\em Phys. Rev. D}
  {\bfseries 68} (2003) 024029},
  \href{http://arxiv.org/abs/hep-th/0303114}{{\ttfamily arXiv:hep-th/0303114}}.

\bibitem{Han1999}
T.~Han, J.~D. Lykken, and R.-J. Zhang, ``Kaluza-klein states from large extra
  dimensions,'' \href{http://dx.doi.org/10.1103/physrevd.59.105006}{{\em
  Physical Review D} {\bfseries 59} no.~10, (Mar, 1999) }.
  \url{https://doi.org/10.1103\%2Fphysrevd.59.105006}.

\bibitem{Randall:1999ee}
L.~Randall and R.~Sundrum, ``{A Large mass hierarchy from a small extra
  dimension},'' \href{http://dx.doi.org/10.1103/PhysRevLett.83.3370}{{\em Phys.
  Rev. Lett.} {\bfseries 83} (1999) 3370--3373},
  \href{http://arxiv.org/abs/hep-ph/9905221}{{\ttfamily arXiv:hep-ph/9905221}}.

\bibitem{Randall:1999vf}
L.~Randall and R.~Sundrum, ``{An Alternative to compactification},''
  \href{http://dx.doi.org/10.1103/PhysRevLett.83.4690}{{\em Phys. Rev. Lett.}
  {\bfseries 83} (1999) 4690--4693},
  \href{http://arxiv.org/abs/hep-th/9906064}{{\ttfamily arXiv:hep-th/9906064}}.

\bibitem{Chivukula:2020hvi}
R.~S. Chivukula, D.~Foren, K.~A. Mohan, D.~Sengupta, and E.~H. Simmons,
  ``{Massive Spin-2 Scattering Amplitudes in Extra-Dimensional Theories},''
  \href{http://dx.doi.org/10.1103/PhysRevD.101.075013}{{\em Phys. Rev. D}
  {\bfseries 101} no.~7, (2020) 075013},
  \href{http://arxiv.org/abs/2002.12458}{{\ttfamily arXiv:2002.12458
  [hep-ph]}}.

\bibitem{SekharChivukula:2019qih}
R.~Sekhar~Chivukula, D.~Foren, K.~A. Mohan, D.~Sengupta, and E.~H. Simmons,
  ``{Sum Rules for Massive Spin-2 Kaluza-Klein Elastic Scattering
  Amplitudes},'' \href{http://dx.doi.org/10.1103/PhysRevD.100.115033}{{\em
  Phys. Rev. D} {\bfseries 100} no.~11, (2019) 115033},
  \href{http://arxiv.org/abs/1910.06159}{{\ttfamily arXiv:1910.06159
  [hep-ph]}}.

\bibitem{SekharChivukula:2019yul}
R.~Sekhar~Chivukula, D.~Foren, K.~A. Mohan, D.~Sengupta, and E.~H. Simmons,
  ``{Scattering amplitudes of massive spin-2 Kaluza-Klein states grow only as
  ${\cal O}(s)$},'' \href{http://dx.doi.org/10.1103/PhysRevD.101.055013}{{\em
  Phys. Rev. D} {\bfseries 101} no.~5, (2020) 055013},
  \href{http://arxiv.org/abs/1906.11098}{{\ttfamily arXiv:1906.11098
  [hep-ph]}}.

\bibitem{Hang:2021fmp}
Y.-F. Hang and H.-J. He, ``{Structure of Kaluza-Klein graviton scattering
  amplitudes from the gravitational equivalence theorem and double copy},''
  \href{http://dx.doi.org/10.1103/PhysRevD.105.084005}{{\em Phys. Rev. D}
  {\bfseries 105} no.~8, (2022) 084005},
  \href{http://arxiv.org/abs/2106.04568}{{\ttfamily arXiv:2106.04568
  [hep-th]}}.

\bibitem{Hang:2022rjp}
Y.-F. Hang and H.-J. He, ``{Gravitational Equivalence Theorem and Double-Copy
  for Kaluza-Klein Graviton Scattering Amplitudes},''
  \href{http://dx.doi.org/10.34133/2022/9860945}{{\em Research} {\bfseries
  2022} (2022) 9860945}, \href{http://arxiv.org/abs/2207.11214}{{\ttfamily
  arXiv:2207.11214 [hep-th]}}.

\bibitem{Chivukula:2022tla}
R.~S. Chivukula, D.~Foren, K.~A. Mohan, D.~Sengupta, and E.~H. Simmons,
  ``{Spin-2 Kaluza-Klein scattering in a stabilized warped background},''
  \href{http://dx.doi.org/10.1103/PhysRevD.107.035015}{{\em Phys. Rev. D}
  {\bfseries 107} no.~3, (2023) 035015},
  \href{http://arxiv.org/abs/2206.10628}{{\ttfamily arXiv:2206.10628
  [hep-ph]}}.

\bibitem{Chivukula:2021xod}
R.~S. Chivukula, D.~Foren, K.~A. Mohan, D.~Sengupta, and E.~H. Simmons,
  ``{Spin-2 Kaluza-Klein mode scattering in models with a massive radion},''
  \href{http://dx.doi.org/10.1103/PhysRevD.103.095024}{{\em Phys. Rev. D}
  {\bfseries 103} no.~9, (2021) 095024},
  \href{http://arxiv.org/abs/2104.08169}{{\ttfamily arXiv:2104.08169
  [hep-ph]}}.

\bibitem{deGiorgi:2020qlg}
A.~de~Giorgi and S.~Vogl, ``{Unitarity in KK-graviton production: A case study
  in warped extra-dimensions},''
  \href{http://dx.doi.org/10.1007/JHEP04(2021)143}{{\em JHEP} {\bfseries 04}
  (2021) 143}, \href{http://arxiv.org/abs/2012.09672}{{\ttfamily
  arXiv:2012.09672 [hep-ph]}}.

\bibitem{Giudice1999}
G.~F. Giudice, R.~Rattazzi, and J.~D. Wells, ``Quantum gravity and extra
  dimensions at high-energy colliders,''
  \href{http://dx.doi.org/10.1016/s0550-3213(99)00044-9}{{\em Nuclear Physics
  B} {\bfseries 544} no.~1-2, (Apr, 1999) 3--38}.
  \url{https://doi.org/10.1016\%2Fs0550-3213\%2899\%2900044-9}.

\bibitem{Bernal:2018qlk}
N.~Bernal, M.~Dutra, Y.~Mambrini, K.~Olive, M.~Peloso, and M.~Pierre, ``{Spin-2
  Portal Dark Matter},''
  \href{http://dx.doi.org/10.1103/PhysRevD.97.115020}{{\em Phys. Rev. D}
  {\bfseries 97} no.~11, (2018) 115020},
  \href{http://arxiv.org/abs/1803.01866}{{\ttfamily arXiv:1803.01866
  [hep-ph]}}.

\bibitem{Bernal:2020fvw}
N.~Bernal, A.~Donini, M.~G. Folgado, and N.~Rius, ``{Kaluza-Klein FIMP Dark
  Matter in Warped Extra-Dimensions},''
  \href{http://dx.doi.org/10.1007/JHEP09(2020)142}{{\em JHEP} {\bfseries 09}
  (2020) 142}, \href{http://arxiv.org/abs/2004.14403}{{\ttfamily
  arXiv:2004.14403 [hep-ph]}}.

\bibitem{Carrillo-Monteverde:2018phy}
A.~Carrillo-Monteverde, Y.-J. Kang, H.~M. Lee, M.~Park, and V.~Sanz, ``{Dark
  Matter Direct Detection from new interactions in models with spin-two
  mediators},'' \href{http://dx.doi.org/10.1007/JHEP06(2018)037}{{\em JHEP}
  {\bfseries 06} (2018) 037}, \href{http://arxiv.org/abs/1803.02144}{{\ttfamily
  arXiv:1803.02144 [hep-ph]}}.

\bibitem{Voronchikhin:2023znz}
I.~V. Voronchikhin and D.~V. Kirpichnikov, ``{The resonant probing spin-0 and
  spin-2 dark matter mediators with fixed target experiments},''
  \href{http://arxiv.org/abs/2304.14052}{{\ttfamily arXiv:2304.14052
  [hep-ph]}}.

\bibitem{Voronchikhin:2022rwc}
I.~V. Voronchikhin and D.~V. Kirpichnikov, ``{Probing hidden spin-2 mediator of
  dark matter with NA64e, LDMX, NA64\ensuremath{\mu}, and M3},''
  \href{http://dx.doi.org/10.1103/PhysRevD.106.115041}{{\em Phys. Rev. D}
  {\bfseries 106} no.~11, (2022) 115041},
  \href{http://arxiv.org/abs/2210.00751}{{\ttfamily arXiv:2210.00751
  [hep-ph]}}.

\bibitem{Jodlowski:2023yne}
K.~Jod\l{}owski, ``{Looking forward to photon-coupled long-lived particles I:
  massive spin-2 portal},'' \href{http://arxiv.org/abs/2305.05710}{{\ttfamily
  arXiv:2305.05710 [hep-ph]}}.

\bibitem{Bjerrum-Bohr:2014lea}
N.~E.~J. Bjerrum-Bohr, B.~R. Holstein, L.~Plant\'e, and P.~Vanhove,
  ``{Graviton-Photon Scattering},''
  \href{http://dx.doi.org/10.1103/PhysRevD.91.064008}{{\em Phys. Rev. D}
  {\bfseries 91} no.~6, (2015) 064008},
  \href{http://arxiv.org/abs/1410.4148}{{\ttfamily arXiv:1410.4148 [gr-qc]}}.

\bibitem{Mathematica}
W.~R. Inc., ``Mathematica, {V}ersion 13.2.''
\newblock \url{https://www.wolfram.com/mathematica}. Champaign, IL, 2022.

\bibitem{Williams__2022}
A.~G. Williams, \href{http://dx.doi.org/10.1017/9781108585286}{{\em
  Introduction to Quantum Field Theory: Classical Mechanics to Gauge Field
  Theories}}.
\newblock Cambridge University Press, 2022.

\bibitem{Falkowski:2020mjq}
A.~Falkowski and G.~Isabella, ``{Matter coupling in massive gravity},''
  \href{http://dx.doi.org/10.1007/JHEP04(2020)014}{{\em JHEP} {\bfseries 04}
  (2020) 014}, \href{http://arxiv.org/abs/2001.06800}{{\ttfamily
  arXiv:2001.06800 [hep-th]}}.

\end{thebibliography}\endgroup

\appendix
\clearpage
\section{Supplementary Material}
\subsection{Feynman Rules}
The Feynman rules for the various vertices are demonstrated below and agree with \cite{Han1999}. In these diagrams, momentum flows left to right:
\begin{figure}[H]
    \centering
    \includegraphics[width=0.25\textwidth]{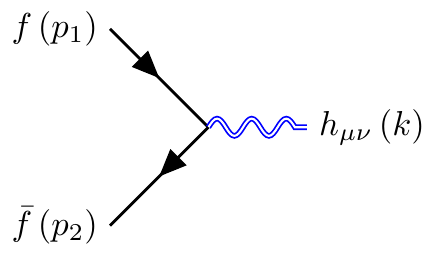}
    \label{fig_ffh}
    \begin{align}
        & = -\dfrac{i\kappa}{8} \bigg[\left(p_{1} - p_{2}\right)_{\mu}\gamma_{\nu} + \left(p_{1} - p_{2}\right)_{\nu}\gamma_{\mu}\nonumber \\
        &\qquad\qquad\qquad- 2\eta_{\mu\nu}\left(\slashed{p}_{1} - \slashed{p}_{2} - 2m_{f}\right)\bigg],
    \end{align}
    \includegraphics[width=0.25\textwidth]{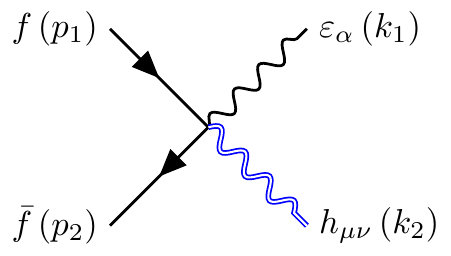}
    \label{fig_ffph}
    \begin{align}
        & = \dfrac{i\kappa g_{f}e}{4}\bigg[\gamma_{\mu}\eta_{\nu\alpha} + \gamma_{\nu}\eta_{\mu\alpha} - 2\eta_{\mu\nu}\gamma_{\alpha}\bigg],
    \end{align}
    \includegraphics[width=0.25\textwidth]{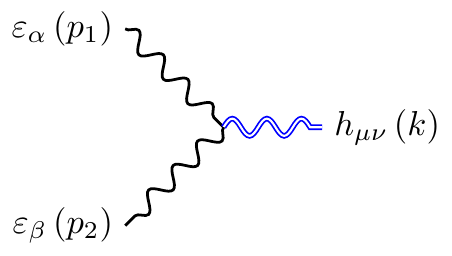}
    \label{fig_pph}
    \begin{align}
        &= -\dfrac{i\kappa}{2}\left[W_{\mu\nu\alpha\beta}\left(-p_{1}, -p_{2}, \xi\right) + W_{\nu\mu\alpha\beta}\left(-p_{1}, -p_{2}, \xi\right)\right],
    \end{align}
    \includegraphics[width=0.25\textwidth]{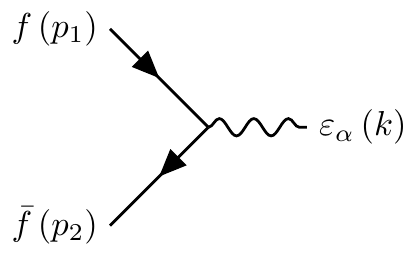}
    \label{fig_ffp}
    \begin{align}
        & = -ig_{f}e\gamma_{\alpha}.
    \end{align}
\end{figure}

\subsection{Cancellations in the Helicity-1 Modes}
A limit-taking process is performed in Mathematica on each channel resulting from the different Feynman diagrams, and we find a definite divergence in the limit as $M_{G}\rightarrow0$ for individual diagrams. The leading divergent terms for the polarization mode $\left(u, \bar{v}, \gamma, G\right) = \left(\uparrow,\uparrow,+1,+1\right)$ are provided in Table \ref{tab:hel1}.

We show that cancellations in the matrix element in the helicity-1 mode are identical to that of the helicity-0 mode. As before, the divergent pieces in the $t$ and $u$ channels are exactly cancelled by the divergent pieces in the $s$-channel and contact term, such that the limit as $M_{G}\to 0$ is regular. As we scan through all polarization combinations, we find that the same pattern follows for each polarization state for the vector modes of the graviton.
\vspace{2ex}
\begin{table}[t]
\begin{tabular}{|p{1cm}||p{6.5cm}|}
    \hline
    \multicolumn{2}{|c|}{Vector Polarization: $\left(u, \bar{v}, \gamma, G\right) = \left(\uparrow,\uparrow,+1,+1\right)$} \\
    \hline
    \hline
    \multicolumn{2}{|c|}{Coefficients of: $\left(\kappa g_{f} e/2\right)\sqrt{s/2}\left(m_{f}/M_{G}\right)$} \\
    \hline
    \hline
    $\mathcal{M}_{t}$ & $\sin^{2}\theta\sqrt{1 - 4m_{f}^{2}/s} + \left(1 - \cos\theta\right)/2$ \\
    \hline 
    $\mathcal{M}_{u}$ & $-\sin^{2}\theta\sqrt{1 - 4m_{f}^{2}/s} - \left(1 + \cos\theta\right)/2$\\
    \hline
    $\mathcal{M}_{s}$ & $2\cos\theta$ \\
    \hline 
    $\mathcal{M}_{c}$ & $-\cos\theta$\\
    \hline
    \hline
    $\sum\mathcal{M}$ & 0\\
    \hline
\end{tabular}
\caption{The cancellations for a helicity-1 external graviton are presented. 
Note that $\mathcal{M}_{t,u,s,c}$ represent the matrix element contributions for the diagrams depicted in Fig. \ref{fd1} }
\label{tab:hel1}
\end{table}

\subsection{Table of Components for Polarizations}
A series of tables containing the leading divergence piece of each polarization combination for the individual diagrams are shown below. For a box in the tables below containing the number $j$, we mean that the leading order term as $M_{G}\to0$ for the matrix element of graviton polarization $\lambda_{G}$ and photon polarization $\gamma$ is proportional to $(M_{G})^{j}$. The choice of fermion spins were inconsequential to the result, and so the number represents the leading divergent piece for all choices of fermion spins. 

In Table \ref{tab:tu}, we highlight that each polarization combination grows as expected by power counting. In Tables \ref{tab:s} and \ref{tab:c}, we find that some polarization combinations vanish identically. We expect the $t$ and $u$-channel divergences to cancel directly in these instances. Interestingly, the contact term is regular in the longitudinal mode but not the vector mode. The divergent contribution here is proportional to $\varepsilon_{0}\left(k_{2}\right)\cdot\varepsilon_{\lambda_{\gamma}}\left(k_{1}\right) = 0$, which evaluates to zero as they are orthogonal in the centre-of-momentum frame. 
\vspace{4ex}
\begin{table}[t]
\centering
\begin{tabular}{|c|c||c||c||c||c||c|}
    \hline
    \multicolumn{7}{|c|}{$\mathcal{M}_{t,u}\sim\mathcal{O}\left(M_{G}\right)$} \\
    \hline
    \hline
    \multicolumn{2}{|c||}{} & \multicolumn{5}{c|}{$\lambda_{G}$} \\
    \cline{3-7}
    \multicolumn{2}{|c||}{} & $-2$ & $-1$ & $0$ & $+1$ & $+2$ \\
    \hline
    \multirow{2}{*}{$\gamma$} & $+1$ & 0 & -1 & -2 & -1 & 0 \\
    \cline{2-7}
    & $-1$ & 0 & -1 & -2 & -1 & 0 \\
    \hline
\end{tabular}
\caption{$t$-Channel and $u$-Channel Divergence Breakdown}
\label{tab:tu}
\end{table}
\vspace{4ex}
\begin{table}[t]
\centering
\begin{tabular}{|c|c||c||c||c||c||c|}
    \hline
    \multicolumn{7}{|c|}{$\mathcal{M}_{s}\sim\mathcal{O}\left(M_{G}\right)$} \\
    \hline
    \hline
    \multicolumn{2}{|c||}{} & \multicolumn{5}{c|}{$\lambda_{G}$} \\
    \cline{3-7}
    \multicolumn{2}{|c||}{} & $-2$ & $-1$ & $0$ & $+1$ & $+2$ \\
    \hline
    \multirow{2}{*}{$\gamma$} & $+1$ & 0 & 0 & -2 & -1 & 0 \\
    \cline{2-7}
    & $-1$ & 0 & -1 & -2 & 0 & 0 \\
    \hline
\end{tabular}
\caption{$s$-Channel Divergence Breakdown}
\label{tab:s}
\end{table}
\vspace{4ex}
\begin{table}[H]
\centering
\begin{tabular}{|c|c||c||c||c||c||c|}
    \hline
    \multicolumn{7}{|c|}{$\mathcal{M}_{c}\sim\mathcal{O}\left(M_{G}\right)$} \\
    \hline
    \hline
    \multicolumn{2}{|c||}{} & \multicolumn{5}{c|}{$\lambda_{G}$} \\
    \cline{3-7}
    \multicolumn{2}{|c||}{} & $-2$ & $-1$ & $0$ & $+1$ & $+2$ \\
    \hline
    \multirow{2}{*}{$\gamma$} & $+1$ & 0 & 0 & 0 & -1 & 0 \\
    \cline{2-7}
    & $-1$ & 0 & -1 & 0 & 0 & 0 \\
    \hline
\end{tabular}
\caption{Contact Term Divergence Breakdown}
\label{tab:c}
\end{table}
\end{document}